\documentclass[runningheads,citeauthoryear]{apinv}
\usepackage{epsfig,cite,graphics}
\usepackage{marvosym} %For astronomical symbols % % %
\usepackage{multirow}
\usepackage[utf8]{inputenc}
\usepackage{float}

\begin{document}

\title{Estimation of radial velocities of BHB stars}
\titlerunning{}
\author{Tahereh Ramezani\inst{1}, Ernst Paunzen\inst{1}, Caiyun Xia\inst{1}, Kate{\v r}ina Pivonkov{\'a}\inst{1}, and Prapti Mondal\inst{1}}
\authorrunning{T.~Ramezani, E.~Paunzen, C.~Xia, K.~Pivonkov{\'a}, and P.~Mondal}
\tocauthor{Tahereh Ramezani, Ernst Paunzen, Caiyun Xia, Kate{\v r}ina Pivonkov{\'a}, and Prapti Mondal} 
% Command tocautor{} is used by the Latex to give author names 
% to the Contents of the volume (automatically generated)
\institute{Department of Theoretical Physics and Astrophysics, Masaryk University, Kotl\'a\v{r}sk\'a 2, CZ-611\,37 Brno, Czechia
	  \newline
	\email{514009@mail.muni.cz}    }
\papertype{Submitted on 11 Novemver 2023; Accepted on 05 April 2024}	
% Papertype can be "Research report", "Review", "Invited lecture", "Conference talk", 
% "Conference poster", "Lecture at scientific seminar", "Summary of dissertation",  etc.
\maketitle

\pagenumbering{gobble} % suppress page-number

\begin{abstract}
We studied blue horizontal branch stars (BHBs), and calculated their radial velocities. Spectra of these stars have been obtained with moderate signal-to-noise ratio for five blue horizontal-branch stars using the 2 meter telescope and Echelle Spectrograph in Ond\v{r}ejov observatory, Czech republic.

%We find that the average radial velocity of our BHB star sample in the direction of Galactic rotation is $\ V_{\rm 0} =~ $-$81 \pm 0.3 ~ {\rm km/s}$.
   
\end{abstract}
  

\section*{Introduction}

Blue horizontal branch (BHB) stars are hot helium burning stars found at
the blue edge of the horizontal branch. BHBs are quite often used to study the properties of the galactic halo (Utkin \& Dambis 2020; Kinman et al. 2000; Vickers et al. 2021). One can think of these stars as a kind of standard candle due to their relatively constant absolute magnitude, with value around zero, which means we are able to predict their distances with high precision, thus also the distance to the Galactic centre (Utkin \& Dambis 2020). In the catalogue of Vickers et al. (2021), the absolute magnitude of almost 14\,000 stars was predicted to be $M_V~\approx{0.31}$~mag.

Subdwarf stars of spectral type B (sdBs) are extremely hot stars found at the end of the horizontal branch, related to the blue horizontal branch (BHB) stars, which are quite often used to trace the galactic halo (Kinman et al. 2000).
Many scenarios to explain the evolution of sdBs have been proposed during last decades. Some such scenarios are strong mass loss rate occurring in a close binary system (Bailer-Jones et al. 2018), mergering of a close binary white dwarfs (Bernhard et al. 2015) or a merging of a red giant and a low mass main sequence star during a common envelope phase leading to formation of a rapidly rotating hot subdwarf star, proposed by Politano et al. (2008).

The stars of type sdB show similar spectral characteristics to the main sequence stars of the same spectral type. A small difference between sdBs and main sequence stars of the same spectral type is that sdB stars have weaker helium lines, are less luminous and have stronger hydrogen Balmer lines caused by their high surface gravity $(\log g = 5.0-6.0)$, while at the same time they are quite compact objects $(R_{\rm sdB}=0.1-0.3~R_\odot)$ with very thin hydrogen envelope and helium core (Geier \& Heber 2012). Irrgang et al. (2021) assumed a typical mass for BHB stars as $\ 0.5\pm 0.1~M_\odot$.

According to Philip et al. (1990), almost no rotational broadening in high resolution spectra is visible and an appropriate Str{\"o}mgren $(b-y)$ and ${\rm c}_1$ criteria must be fulfilled for BHB stars. %Must be changed 1995 ...
They found out a high rotation velocity for HB stars and low rotation for hotter stars.

\section{Observations}

The observations were obtained during a few nights in September 2021 (8th and 10th) with the Perek 2 meter telescope located in Ond{\v r}ejov, Czech republic. 

The Perek 2 meter telescope has a resolving power of R$\sim$13\,000 at H$\alpha$. Additionally, the telescope is equipped with two instruments, a single-order spectrograph and an Echelle spectrograph.

%The Echelle spectrograph has a resolving power of R=51600 at 5000 \AA (R$\sim$ 40000 in $H\alpha$) and spectral sampling is 2.4 \AA/mm. Therefore, the resolving power of the Perek 2m telescope depends on the instrument used, with the Echelle spectrograph having a higher resolving power than the single-order spectrograph.

Those objects have been observed spectroscopically with the Echelle spectrograph connected to the Perek 2 meter telescope.
Each spectrum was processed using the oesred.cl script designed for work in IRAF by Cabezas et al. (2023).
Estimation of radial velocities was done using the SpaS software designed for plotting and following analysis of 1D spectra.

Spectra of all five BHB stars investigated are shown in the Figure 1. 

%%%%% fig1
\begin{figure}[htp]  %fig1
    \centering
    \includegraphics[width=12cm]{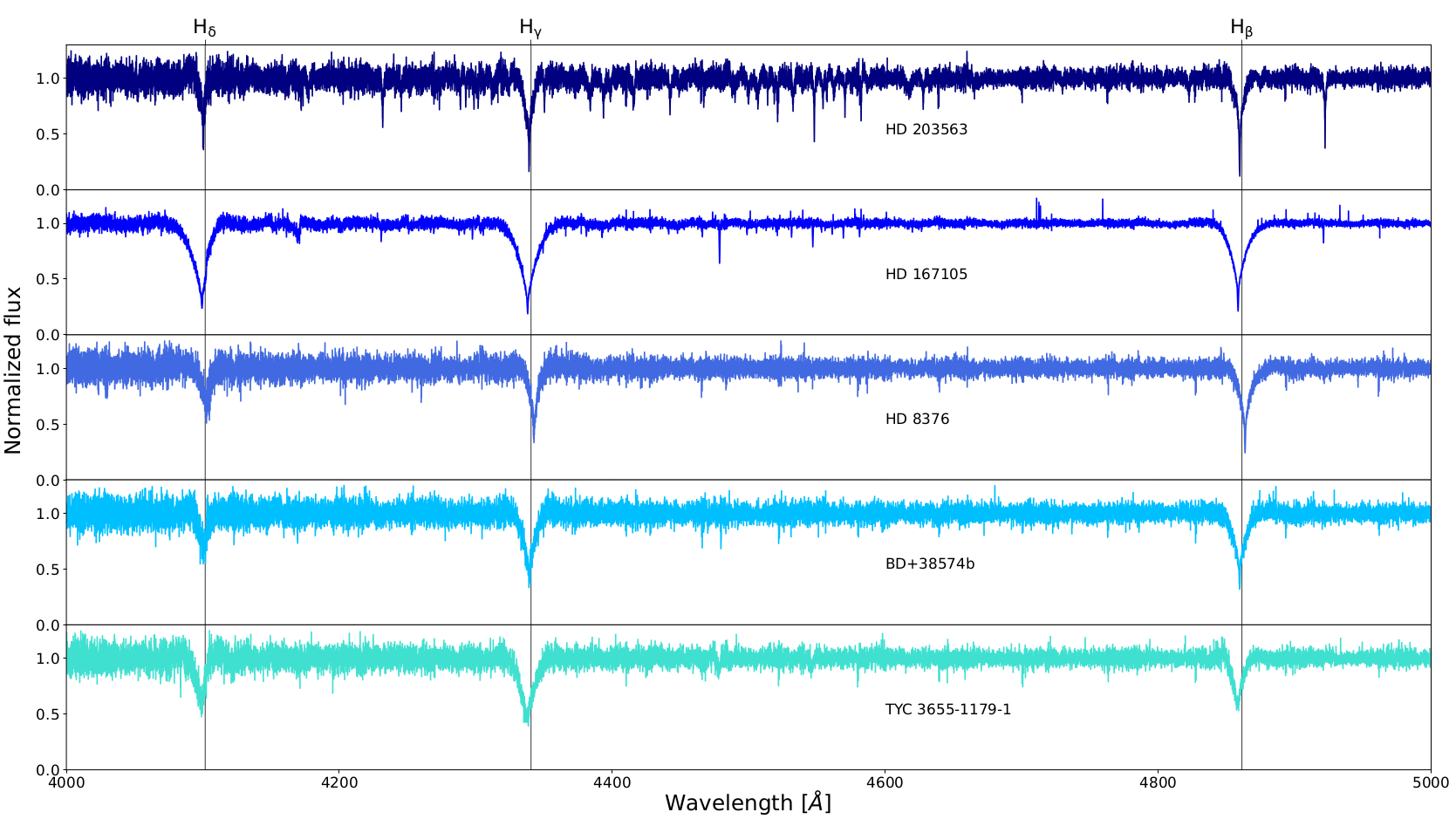}
    \caption{Spectra of the five BHB stars studied.}
\end{figure}

The journal of observations of our stars is given in Table 1.

\begin{table}[H]
\centering
\caption{List of stars with nights of observation (first epoch and second epoch), number of observations per star, and the exposure time of each spectrum.}
\label{table3}   
\begin{tabular}{|c|c|c|c|c|}
\hline
\multirow{2}{*}{star ID} &  \multirow{2}{*}{first epoch} & \multirow{2}{*}{second epoch} & \multirow{2}{*} {number of observations} &  \multirow{2}{*}{exposure time [s]} \\
 & & & & \\
\hline
TYC3655-1179-1 & 08.09.2021 & 10.09.2021 & 2 & 5200 \\
\hline
HD 8376 & 08.09.2021 & 10.09.2021 & 2 &  3600 \\
\hline
BD+38574b & 08.09.2021 & 10.09.2021 & 2 &  5400 \\
\hline
HD 167105 & 08.09.2021 & 10.09.2021 & 2 &  3600 \\
\hline
HD 203563 & 08.09.2021 & 10.09.2021 & 2 &  1039 \\
\hline
\end{tabular}
\end{table}

\section{Data analysis}

We got the reddening from Schlafly \& Finkbeiner (2011), distance from Bailer-Jones et al. (2021), and coordinates from Gaia EDR3 (Gaia Collaboration 2020) for each object and put them in the Table~2.

We estimated radial velocities for the first epoch and the second epoch with their standard deviations for each epoch and put them in Table~3.

\begin{table}[h!]
\centering
\caption{List of stars for which we got reddening from Schlafly \& Finkbeiner (2011), and distance from Bailer-Jones et al. (2021). We got Longitude and Latitude values from Gaia EDR3 (Gaia Collaboration 2020).}
\label{table1}    
\begin{tabular}{|c|c|c|c|c|c|c|}
\hline
\multirow{2}{*}{star ID} & \multirow{2}{*}{RA} & \multirow{2}{*}{DEC} & \multirow{2}{*}{Longitude} & \multirow{2}{*}{Latitude} & \multirow{2}{*} {{E(B-V)} [mag]} & \multirow{2}{*} {{Distance} [pc]}\\
 & & & & & & \\ 
\hline
TYC3655-1179-1 & 00 44 43.32 & +53 10 39.22 & 121.911 & -9.681 & 0.156 & 739 $\pm$~ 9 \\
\hline
HD 8376 & 01 23 28.28 & +31 47 12.26 & 130.839 & -30.593 & 0.045 & 624 $\pm$~ 6 \\
\hline
BD+38574b & 02 51 44.00  & +38 53 47.95  & 147.178 & -18.246 & 0.052 & 652 $\pm$~ 6 \\
\hline
HD 167105 & 18 11 06.30  & +50 47 32.41 & 78.694 & +26.861 & 0.062 & 405 $\pm$~ 2 \\
\hline
HD 203563 & 21 22 59.96 & +02 14 37.29 & 54.686 & -31.892 & 0.026 & 441 $\pm$~ 5 \\
\hline
\end{tabular}
\end{table}

\begin{table}[h!]
\centering
\caption{List of stars for which we estimated radial velocities for the first epoch (second column) and the second epoch (fourth column) with their standard deviations for each epoch.}
\label{table2}   
\begin{tabular}{|c|c|c|c|c|}
\hline
\multirow{2}{*}{star ID} & \multirow{2}{*} {$v_{\rm rad}$ [km\,s$^{-1}$]} & \multirow{2}{*} {stddev} & \multirow{2}{*} {$v_{\rm rad}$ [km\,s$^{-1}$]} & \multirow{2}{*} {stddev} \\
 & & & &\\
\hline
TYC3655-1179-1 & $-$189 & 0.10 & $-$188 & 0.18 \\
\hline
HD 8376 & $+$144 & 0.08 & $+$145 & 0.23 \\
\hline
BD+38574b & $-$101 & 0.08 & $-$101 & 0.32 \\
\hline
HD 167105 & $-$157 & 0.03 & $-$169 & 0.03 \\
\hline
HD 203563 & $-$101 & 0.05 & $-$96 & 0.10 \\
\hline
\end{tabular}
\end{table}

The Diffuse Interstellar Bands (DIBs) can be as weak as a couple of percent, so this is simply not enough to detect them with certainty. ESO archive has a spectrum of HD 203563, there it has an intensity of 5780\AA\ DIB of around 3\%.

We found our targets using the TOPCAT software. Then we had a night spectroscopy observation where we observed our targets all night and wrote down the start time, which was 21:00, the end time and the exposure time. Then we reduced our data in IRAF by the package oesred. Then we changed the format from .fit to .spec. After that we found the radial velocity of H$\rm _\beta$, H$\rm _\gamma$ and H$\rm _\delta$ and fitted them, which they had convenient fits; also calculated signal to noise which are in the Figures 2 to 6.

%%%%%%% fig2
\begin{figure}[H]
\centering
\includegraphics[width=8cm]{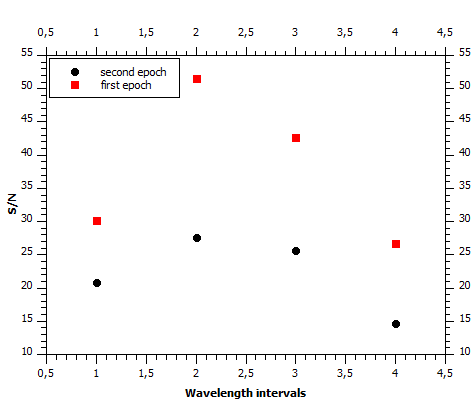}
\caption{%(TYC 3655-1179-1)
The dependence of S/N to wavelength for two different epochs of the same object TYC 3655-1179-1. X axis represents intervals of wavelength: 1. interval is 4000-5000 \AA, 2. interval is 5000-6000 \AA, 3. interval is 6000-7000 \AA, 4. interval is 7000-8000 \AA.}
\label{tyc3655}
\end{figure}

%%%%%%% fig3
\begin{figure}[H]
\centering
\includegraphics[width=8cm]{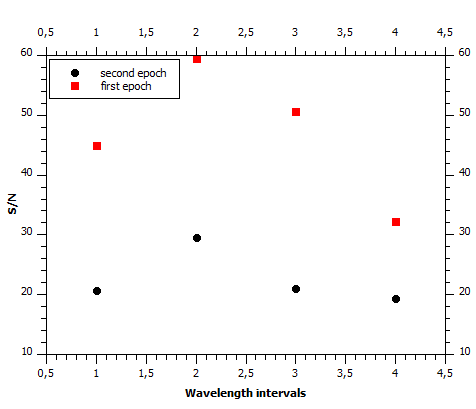}
\caption{%(HD 8376)
The dependence of S/N to wavelength for two different epochs of the same object HD 8376. X axis represents intervals of wavelength: 1. interval is 4000-5000 \AA, 2. interval is 5000-6000 \AA, 3. interval is 6000-7000 \AA, 4. interval is 7000-8000 \AA.}
\label{hd8376}
\end{figure}

%%%%%%% fig4
\begin{figure}[H]
\centering
\includegraphics[width=8cm]{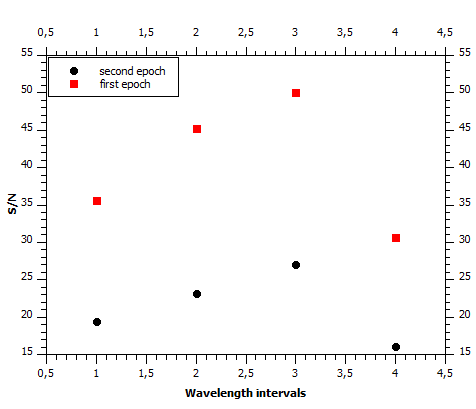}
\caption{%(HD 17708)
The dependence of S/N to wavelength for two different epochs of the same object BD+38574b. X axis represents intervals of wavelength: 1. interval is 4000-5000 \AA, 2. interval is 5000-6000 \AA, 3. interval is 6000-7000 \AA, 4. interval is 7000-8000 \AA.}
\label{hd17708}
\end{figure}

%%%%%%% fig5
\begin{figure}[H]
\centering
\includegraphics[width=8cm]{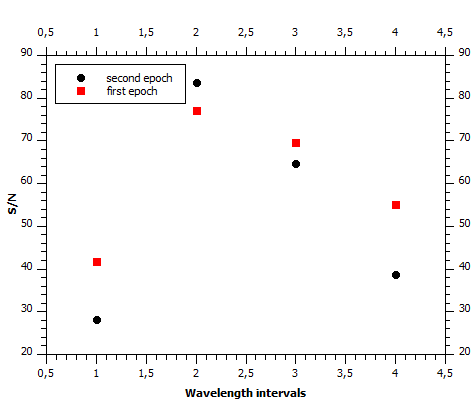}
\caption{%(HD 167105)
The dependence of S/N to wavelength for two different epochs of the same object HD 167105. X axis represents intervals of wavelength: 1. interval is 4000-5000 \AA, 2. interval is 5000-6000 \AA, 3. interval is 6000-7000 \AA, 4. interval is 7000-8000 \AA.}
\label{hd167105}
\end{figure}

%%%%%%% fig6
\begin{figure}[H]
\centering
\includegraphics[width=8cm]{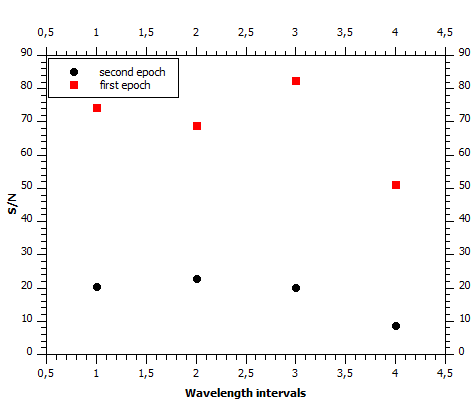}
\caption{%(HD 203563)
The dependence of S/N to wavelength for two different epochs of the same object HD 203563. X axis represents intervals of wavelength: 1. interval is 4000-5000 \AA, 2. interval is 5000-6000 \AA, 3. interval is 6000-7000 \AA, 4. interval is 7000-8000 \AA.}
\label{hd203563}
\end{figure}

%%%%%%% fig7
\begin{figure}[H]
    \centering
    \includegraphics[width=10.2cm]{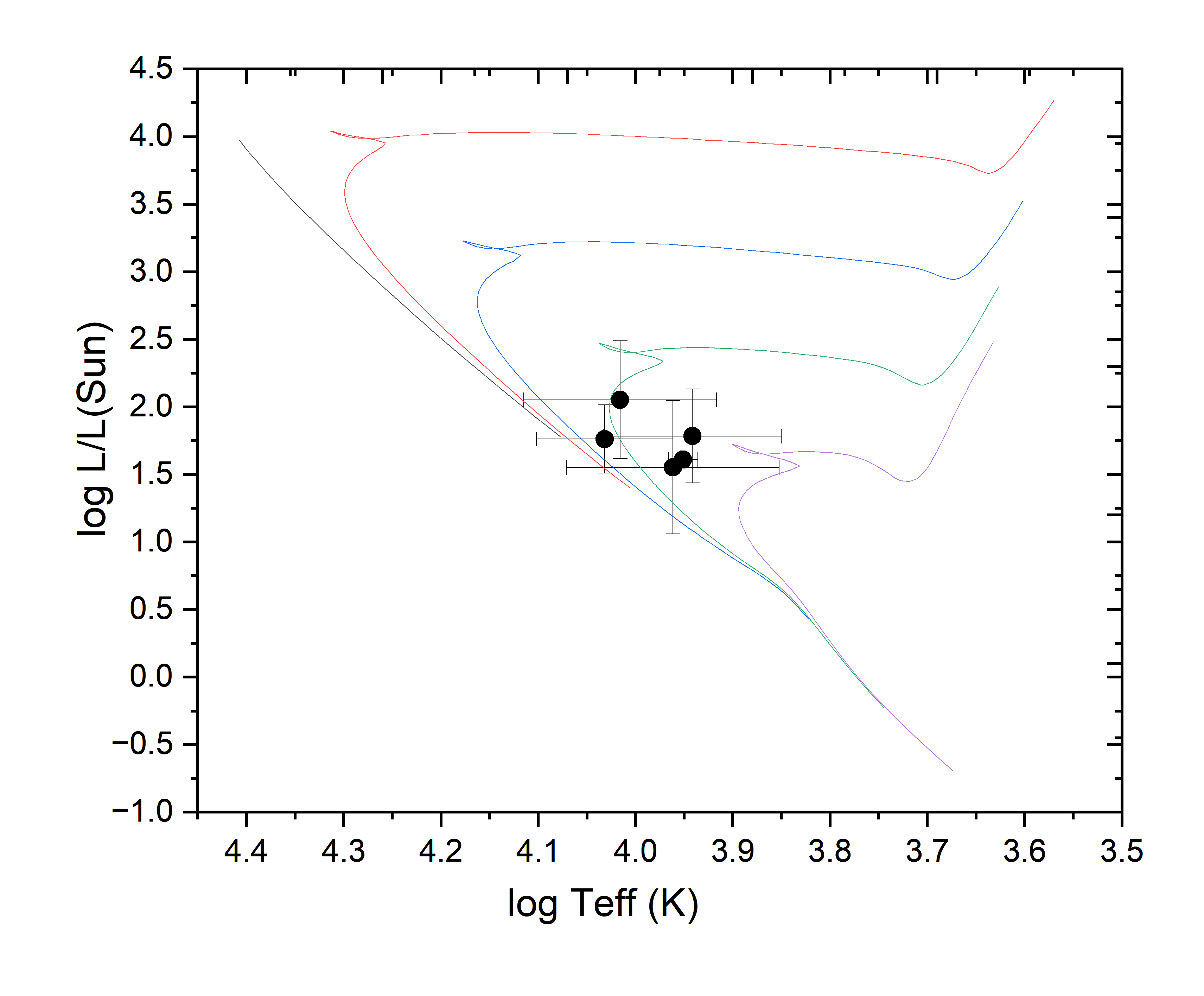}
    \caption{Hertzsprung-Russell diagram of the targets together with isochrones from Bressan et al. (2012).}
\end{figure}

We searched in the catalogues: 
Anders et al. (2022), Paegert et al. (2021), Stassun et al. (2019), McDonald et al. (2017), and from the parameters' values and use of this formula:

%\nonumbering

\begin{equation} 
\log(g /{g_\odot})  =  \log(M / M_\odot)  +  4\log(T_{\rm eff}/{T_{{\rm eff}\odot}}) - \log(L /{L_\odot})
\end{equation}

\noindent from Netopil et al. (2017), we found $\log{T_{\rm eff}}$ and $\log{Lum}$, and plotted the isochrones of the stars which are in the Figure~7.

%%%%%%% fig7

\section{Conclusion}

In our work, we studied five blue horizontal branch stars (BHBs), and estimated their radial velocity using the Echelle spectrograph mounted on the 2 meter telescope situated in Ond\v{r}ejov observatory, Czech Academy of Sciences. We give estimated values of the radial velocities of these stars in Table~\ref{table2}. 

The radial velocity of HD~167105 which was measured by Gontcharov (2006) was $-$172.40 km\,s$^{-1}$, also for BD+38574b which was measured by J\"{o}nsson et al. (2020) was 
$-$94.749496 km\,s$^{-1}$. We see that our measurements are similar to the corresponding values of their measurements. 

We used spectroscopic measurements to determine the stellar parameters of the BHBs. Radial velocities were measured from spectra taken by the 2 meter telescope during the two nights in September 2021. In our samples, we considered only those BHB stars for which the effective temperature $(T_{\rm eff})$ is less than $\sim $10\,000 K.

\section*{Acknowledgements}

This work was supported by Masaryk University, Faculty of Science, and Potsdam university. 
We would like to express our special thanks of gratitude to Jiri Kubat, who gave us the golden opportunity to do this wonderful project, also another supporter, Brankica Kubatova in Ondřejov institute.
From Astronomical Institute of the Czech Academy of Sciences. Thanks to our lecturers, Harry Dawson and Max Pritzkuliet from Potsdam University. We would like to thanks to our colleagues Martin Piecka at the Vienna University, Gabriel Sz{\'a}sz and Jan Janik at the Masaryk University for their guidance.

%\newpage

\end{document}